\documentclass[aps,prb,twocolumn,superscriptaddress,floatfix,longbibliography]{revtex4-1}
\usepackage{amssymb}
\usepackage{graphicx}
\usepackage{dcolumn}
\usepackage{bm}
\usepackage{amsmath}
\usepackage{epstopdf}

\usepackage[colorlinks,linkcolor=magenta,citecolor=blue,urlcolor=blue]{hyperref}

\begin{document}

\title{Mobility edges in $\mathcal{PT}$-symmetric cross-stitch flat band lattices}
\author{Tong Liu}
\thanks{t6tong@njupt.edu.cn}
\affiliation{School of Science, Nanjing University of Posts and Telecommunications, Nanjing 210003, China}
\author{Shujie Cheng}
\affiliation{Department of Physics, Zhejiang Normal University, Jinhua 321004, China}

\date{\today}

\begin{abstract}
We study the cross-stitch flat band lattice with a $\mathcal{PT}$-symmetric on-site potential and  uncover mobility edges with exact solutions. Furthermore, we study the relationship between the $\mathcal{PT}$ symmetry broken point and the localization-delocalization transition point, and verify that mobility edges in this non-Hermitian model is available to signal the $\mathcal{PT}$ symmetry breaking. 

%We uncover mobility edges with exact solutions in two Hermitian and non-Hermitian models. In the Hermitian model, we utilize the self-dual relation to obtain the linear mobility edge which is irrelevant with the quasiperiodic disorder strength, and then verify the conclusions by
%utilizing numerical simulations. In the non-Hermitian model, we study the cross-stitch lattice with a $\mathcal{PT}$-symmetric on-site potential, and identify the exact expression of quadratic mobility edges. Further more, we study the relationship between the $\mathcal{PT}$ symmetry broken point and the localization-delocalization transition point, and verify that mobility edges in this non-Hermitian model not only separate localized from extended states but also indicate the coexistence of complex and real eigenenergies.
\end{abstract}

\maketitle

\section{Introduction}

The unavoidable exchange of the particles, energy and quantum information with surrounding environment forms the so-called open quantum systems \cite{open_1,open_2,open_3,open_4,open_5}. For some insightful considerations, the quantum phenomena of these systems can be well discussed by the effective non-Hermitian Hamiltonians \cite{eff_1,eff_2,eff_3,eff_4,eff_5}. Due to the non-Hermiticity, the eigenvalues of systems become complex, which is the consequence of the non-conservation of possibility. Nevertheless, if systems possess the parity-time ($\mathcal{PT}$) symmetry, they may still have purely real energy spectra, and the presence of the real spectra implies that the gain and loss of systems are balanced \cite{PT_seminal}.
%What needs to be pointed out is that the $\mathcal{PT}$-symmetry the combination of the parity and time symmetry.
Except the particular circumstance \cite{except_1,except_2} that the real spectra appear in systems without
$\mathcal{PT}$-symmetry, we generally name the real-complex transition as the $\mathcal{PT}$-symmetry transition.

Since the seminal work by Bender and Boettcher \cite{PT_seminal}, the study of the $\mathcal{PT}$-symmetry has been much active in the fields of quantum field theories and mathematical physics \cite{PT_seminal_1,PT_seminal_2,PT_seminal_3,PT_seminal_4}, condensed matter physics \cite{cmp_1,cmp_2}, and optical systems \cite{opt_1,opt_2,opt_3,opt_4,opt_5,opt_6}.
Thanks to the progress of the experimental technology, the gain and loss can be engineered controllably , which is conducive to the observation of the $\mathcal{PT}$-symmetry transition \cite{cgl_1,cgl_2,cgl_3,cgl_4,cgl_5,cgl_6,cgl_7,cgl_8,cgl_9}.

Similar to the $\mathcal{PT}$-symmetry, the study of Anderson localization is also a hot field, which is initially uncovered by P. W. Anderson \cite{Anderson}. Anderson localization refers to the breakdown of the diffusion of wave packets due to the disorder. One-dimensional non-interacting systems provide good platforms to study the localization transition. A representative example is the Aubry-Andr\'{e} (AA) model with quasiperiodic on-site potential, which presents the feature of the correlated disorder \cite{AA}.
%In a loose definition, we name the correlated disorder as quasidisorder.
The AA model undergoes a delocalization-localization transition with the increasing strength of the quasiperiodic potential, and the phase transition point can be extracted by the self-dual condition. This localization transition has been observed in the bichromatic optical lattice of ultracold-atom experiments \cite{exp_AA}. During the decades, the AA model has drawn many theoretical and experimental researches~\cite{Laurent1,Laurent2,Laurent3,Laurent4,Laurent5,Laurent6,Billy}.

Recently, the localization transition is explored in some non-Hermitian systems, such as the non-Hermitian Hatano-Nelson model with asymmetric hoppings \cite{HN_1,HN_2,HN_3} and the generalized non-Hermitian AA models \cite{NH_AA_2,NH_AA_3,NH_AA_4,Longhi1,Longhi2}. Gong et.al. presented an intriguing topological explanation about the presence of the localization transition in a non-Hermitian Hatnano-Nelson model \cite{HN_5}. And Schiffer et.al. investigated a generalized AA model with $\mathcal{PT}$-symmetry and uncovered the $\mathcal{PT}$-symmetry protected localization phase \cite{PT_AA}.

Nowadays, the combination of non-Hermiticity and the quasiperiodic potential has gradually intrigued interest in the aspect of non-Hermitian effect on the mobility edge. The physical concept of mobility edges was first proposed by Mott, based on the 3D Anderson model \cite{mott}. The mobility edge refers to a critical energy-level which separating localized from extended states. In the following study, various AA-like models containing mobility edges are discussed, such as slow-varying potentials \cite{slow_1,slow_2}, off-diagonal disorder \cite{off_1,off_2}, long-range hoppings \cite{long-range}, and other generalized quasiperiodic potentials \cite{general_quasi_1,general_quasi_2,general_quasi_3}.
%Moreover, mobility edges are observed in experiments \cite{exp_1,exp_2,exp_3,exp_4,exp_5,exp_6,exp_7}.
Y. Liu et.al. \cite{YLiu} numerically found the simultaneous occurrence of the localization transition and the $\mathcal{PT}$-symmetry breaking. Zeng et.al. \cite{QZeng} demonstrated the correspondence between the winding number and the localization transition, and numerically uncovered mobility edges in the spectrum with or without $\mathcal{PT}$-symmetry. T. Liu et.al. \cite{TLiu} uncovered the existence of the generalized Aubry-Andr\'{e} self-dual symmetry and obtained the exactly analytical mobility edges in non-Hermitian quasicrystals.

However, as far as we know, the influence of non-Hermitian perturbations on the flat bands has not been studied. Flat band lattices~\cite{Flach1,Flach2,Flach3,Flach5,Gneiting} are translationally invariant tight-binding lattices which support at least one dispersionless band in the energy spectrum. Flat band systems have usually been considered as an ideal playground to explore the strong correlation phenomena as a result of the complete quenching of the kinetic energy of electrons. Such as, a nearly flat
band with non-trivial topology was proposed to simulate fractional Chern insulators~\cite{LIU}.
The classification~\cite{Flach1} through compact localized states (CLS) gives a good framework of the properties of flat bands, i.e., the number $U$ of unit cells occupied by a CLS. For the $U = 1$ class, the CLSs form a set of orthogonal and complete bases~\cite{Flach1}, indicating that a single CLS is disentangled from the rest of unit cells, such as the cross-stitch network. However, for generic $U > 1$ classes, the CLSs are not orthogonal to each other in one dimension, such as the sawtooth network~\cite{Flach2}. For exact solvability, in this work, we focus on the study of cross-stitch lattices under non-Hermitian quasiperiodic perturbations.

\section{Model and Mobility edges}
\begin{figure}[t]
  \centering
  \includegraphics[width=0.5\textwidth]{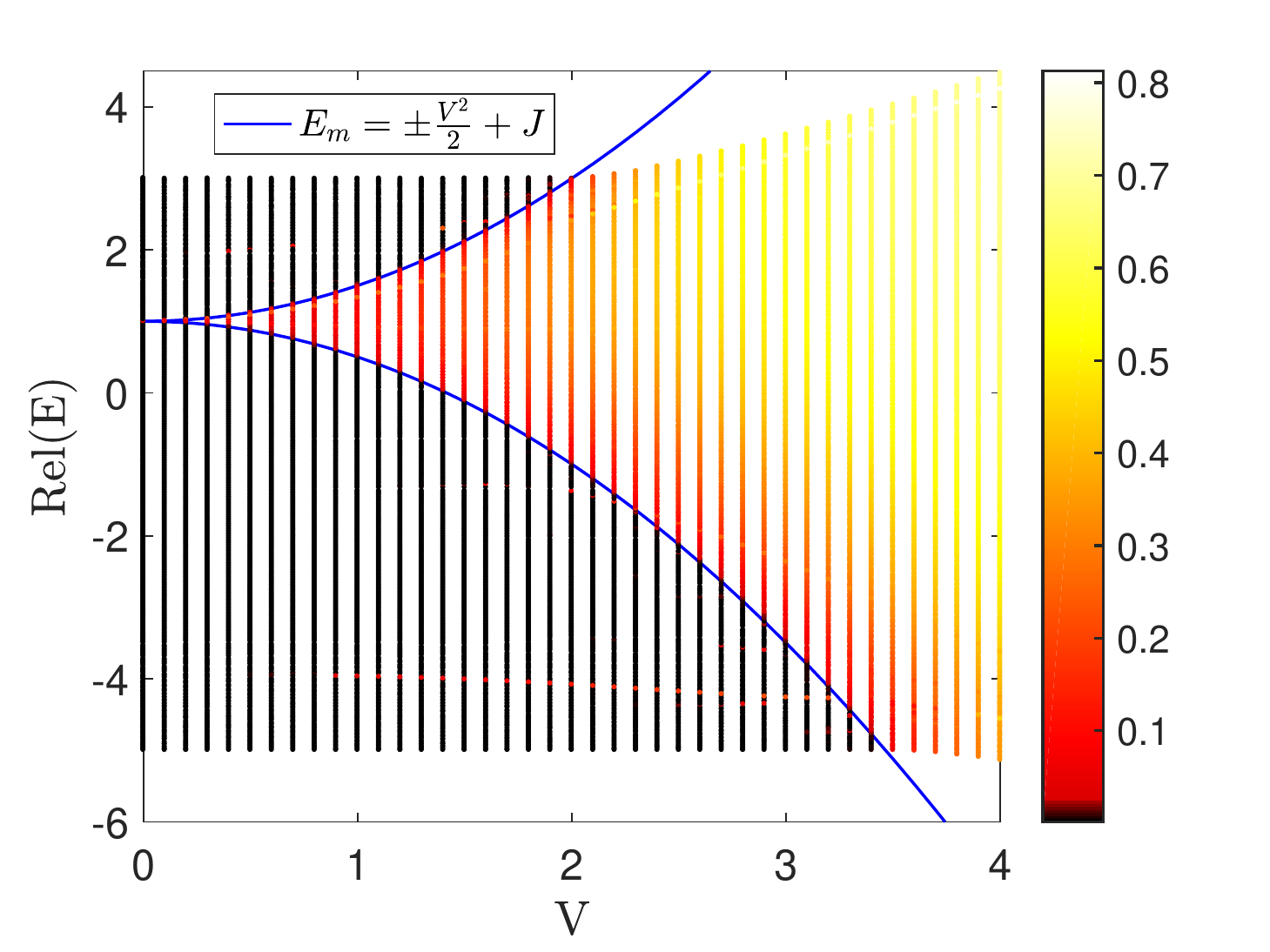}\\
  \caption{(Color online) The real part of eigenvalues of Eq.~(\ref{Eq1}) and IPR as a function of $V$ with the parameter $J=1$. The total number of sites is set to be $L=500$. Different colours of the eigenvalue curves indicate different magnitudes of the IPR of the corresponding wave functions. The black eigenvalue curves denote the extended states, and the bright yellow eigenvalue curves denote the localized states. The blue solid lines represent the boundary between spatially localized and extended states, i.e., the mobility edges $E_m=\pm\frac{V^2}{2}+J$.}
  \label{001}
\end{figure}
We consider a non-Hermitian cross-stitch lattice with the complex on-site potential,
\begin{equation}
\hat{\epsilon}_n\Psi_n - \hat{V}\Psi_n - \hat{T}(\Psi_{n-1} + \Psi_{n+1}) = E\Psi_n\ ,
\label{Eq1}
\end{equation}
with
\begin{equation}
\hat{\epsilon}_n = \left(
\begin{array}{ccc}
\epsilon_n & 0 \\
0 & -\epsilon_n
\end{array} \right),\quad
\hat{V} = \left(
\begin{array}{ccc}
0 & J \\
J & 0
\end{array} \right),\quad
\hat{T} = \left(
\begin{array}{ccc}
1 & 1 \\
1 & 1
\end{array} \right)\;.
\label{Eq2}
\end{equation}
where the inter-cell hopping strength is set to be $1$ and $J$ is the intra-cell hopping strength. In the absence of the potential, $\epsilon_n=0$, there is exactly one flat band $E_{fb} = J$, associated with compact localized states $\Psi_n\equiv(a_n,b_n)^T=(-1,1)^T \delta_{n,n_0}/\sqrt{2}$, and one dispersive band $E(k) = -4 \cos(k) - J$.
$\epsilon_n=V e^{i 2\pi\alpha n}$ is the complex on-site potential.

Applying the local rotations
\begin{equation}
\hat{U} = \frac{1}{\sqrt{2}}\left(
\begin{array}{ccc}
1 & 1 \\
1 & -1
\end{array} \right),\quad
\left(
\begin{array}{cc}
p_n  \\
f_n
\end{array} \right) = \hat{U}\Psi_n \;.
\label{Eq3}
\end{equation}
Eq.~(\ref{Eq1}) becomes
\begin{equation}
\begin{aligned}
&(E+J)p_n=\epsilon_nf_n-2(p_{n-1}+p_{n+1}),\\
&(E-J)f_n=\epsilon_np_n,
\label{Eq4}
\end{aligned}
\end{equation}

According to the method in Ref.\cite{Bodyfelt,Danieli}, by solving for the Fano coordinates $f_n$ in Eq.~(\ref{Eq4}), we obtain a new equation for the dispersive portion,
\begin{equation}
(E+J)p_n=\frac{\epsilon_n^2}{E-J}p_n-2(p_{n-1}+p_{n+1}),
\label{Eq5}
\end{equation}
Recalling $\epsilon_n=V e^{i 2\pi\alpha n}$, Eq.~(\ref{Eq5}) becomes
\begin{equation}
(E+J)p_n=\frac{V^2}{E-J}e^{i 4\pi\alpha n}p_n-2(p_{n-1}+p_{n+1}),
\label{Eq6}
\end{equation}
if we make
\begin{equation}
\begin{aligned}
&\tilde{E}:=E+J,\\
&\tilde{V}:=\frac{V^2}{E-J},
\label{Eq7}
\end{aligned}
\end{equation}
Eq.~(\ref{Eq6}) becomes
\begin{equation}
\tilde{E}p_n=\tilde{V}e^{i 4\pi\alpha n}p_n-2(p_{n-1}+p_{n+1}),
\label{Eq8}
\end{equation}
Numerical and analytical results~\cite{Jazaeri,Longhi1,Longhi2} show that a metal-insulator phase transition arises at the critical point $\tilde{V}=\pm2$ in Eq.~(\ref{Eq8}). Thus an analytic expression is found for the mobility edge,
\begin{equation}
%\frac{V^2}{2(E_m-J)}=1~\Rightarrow~
E_m=\pm\frac{V^2}{2}+J.
\label{Eq9}
\end{equation}

To support the analytical result given above, we now present detailed numerical analysis of Eq.~(\ref{Eq1}). In the disordered system, the localization property of wave functions can be measured by the inverse participation ratio (IPR)~\cite{IPR}. For any given normalized wave function, the corresponding IPR is defined as $\text{IPR} =\sum_{n=1}^{L} \left|\psi_{n}\right|^{4},$
which measures the inverse of the number of sites being occupied by particles. It is well known that the IPR of an extended state scales
like $L^{-1}$ which approaches zero in the thermodynamic limit. However, for a localized state, since only finite number of sites are
occupied, the IPR is finite even in the thermodynamic limit. In Fig.~\ref{001} we show the numerical IPR diagram in the $[{\rm Re}(E),V]$ plane, different colours of the eigenvalue curves indicate different magnitudes of the IPR of the corresponding wave functions. The black eigenvalue curves denote the extended states, and the bright yellow eigenvalue curves denote the localized states. It is clearly demonstrating two mobility edges separating localized from extended states along the blue curves defined by Eq.~(\ref{Eq9}).

\section{Real-complex transition}
\begin{figure}
  \centering
  \includegraphics[width=0.5\textwidth]{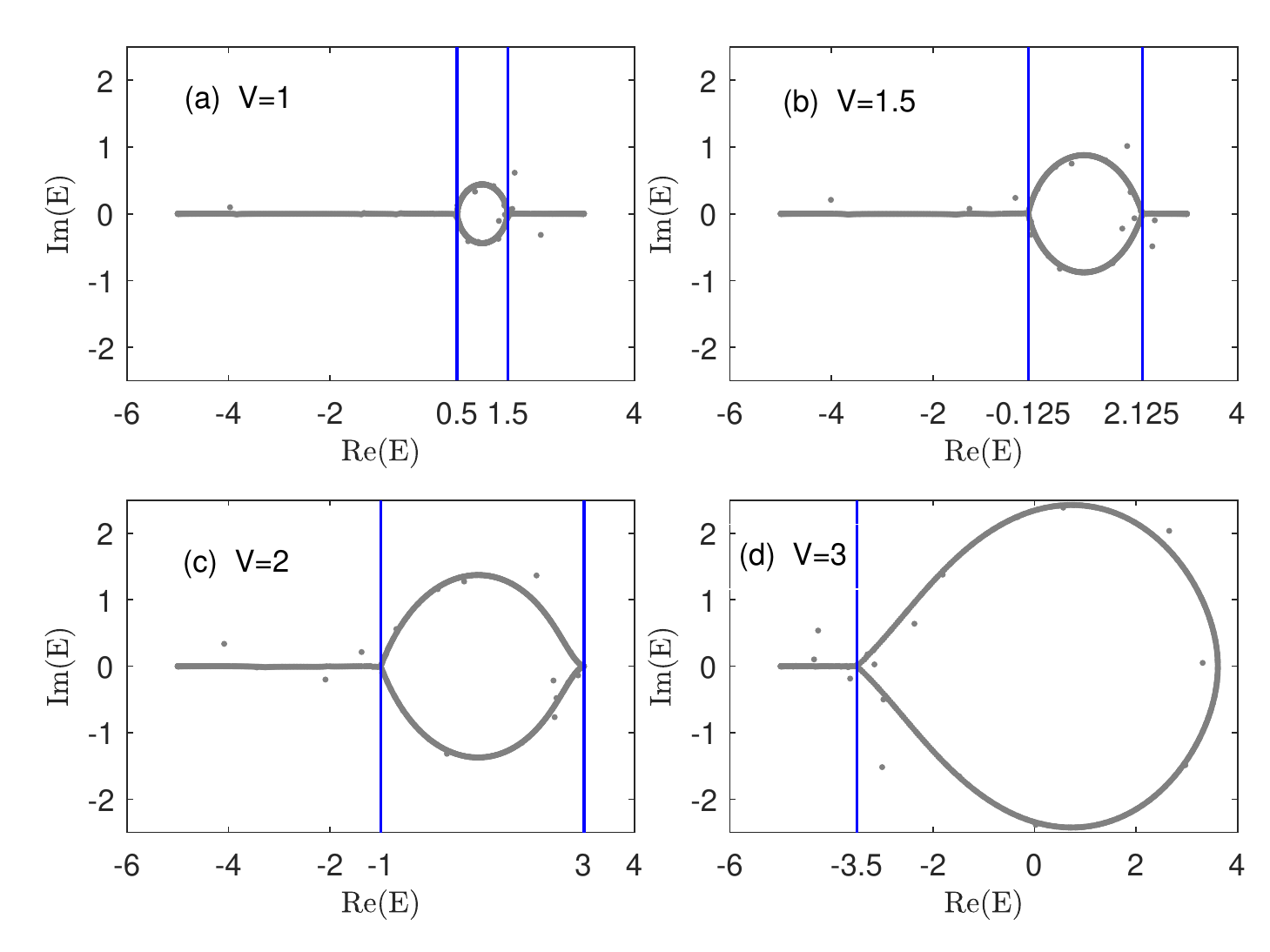}\\
  \caption{Real and imaginary part of eigenvalues for Eq.~(\ref{Eq1}) with the parameter $J=1$ under open boundry conditions. (a) $V=1$, the imaginary part inside the interval $[0.5,1.5]$ is nonzero and eigenvalues form a closed curve, whereas the imaginary part outside the interval $[0.5,1.5]$ is zero and eigenvalues form a line. For other $V$'s, $V=1.5$ (b), $V=2$ (c) and $V=3$ (d), the same real-complex transitions of the spectrum occur.
  The total number of sites is set to be $L=500$. The blue solid lines represent the boundaries between the real and complex energy spectrum, which are in good agreement with the mobility edges $E_m=\pm\frac{V^2}{2}+1$.}
  \label{002}
\end{figure}

By analyzing the energy spectrum, we find that there exists the real-complex transition of spectra.
In Fig.~\ref{002}, we fix the size of the system $L=500$ and plot the eigenvalues of Eq.~(\ref{Eq1}) with various $V$. As the figure~\ref{002}(a) shows, when $V=1$, the eigenvalues outside the interval $[0.5,1.5]$ are real and the system is in the extended phase, whereas those inside the interval $[0.5,1.5]$ are complex  and the system is in the localized phase. The critical energies $E_{min}=0.5$ and $E_{max}=1.5$ are exactly corresponding to the mobility edges $E_m=\pm\frac{V^2}{2}+1$. The results of $V=1.5$, $V=2$ and $V=3$ are also the same, as shown in Fig.~\ref{002}(b), (c) and (d).
Therefore, for each potential strength $V$, we always find the separation of real and imaginary part of the eigenvalues consistent with the exact solution from Eq.~(\ref{Eq9}). The complex energy is accompanied with the localized state, whereas the real energy is accompanied with the extended state. We have also checked other combinations of parameters and get the same results as expected. Consequently, we find a perfect correspondence between the real-complex transition and analytical mobility edge energy.

\section{Summary}
In summary, we have studied the extended and localized phases and investigated the real-complex transition of the cross-stitch flat band lattice subject to the non-Hermitian quasiperiodic potentials.
Firstly, we decouple the cross-stitch lattice and obtain the analytic form of mobility edges in the spectrum. By diagonalizing the Hamiltonian, we numerically obtain the eigenvalues and wave functions. The numerical results clearly show the existence of mobility edges and are in excellent agreement with the theoretical predictions by analysing the inverse participation ratio.
Furthermore, by analysing the energy spectrum, we demonstrate that mobility edges in non-Hermitian potentials not only separate localized from extended states but also indicate the coexistence of real and complex eigenvalues.
Our finding the study of the non-Hermitian mobility edges and the real-complex transition, and in the view of $\mathcal{PT}$ symmetry, the localization transition occurs companied by the $\mathcal{PT}$ symmetry breaking.

\begin{acknowledgments}
T. L.~acknowledges the Natural Science Foundation of Jiangsu Province (Grant No.~BK20200737) and NUPTSF (Grant No.~NY220090 and No.~NY220208).
\end{acknowledgments}


\begin{thebibliography}{10}
% open system and effective non-Hermitian Hamiltonian
\bibitem{open_1} V. V. Sokolov and V. G. Zelevinsky, On a statistical theory of overlapping resonances, Phys. Lett. B \textbf{202}, 10 (1988).
\bibitem{open_2} I. Rotter, A continuum shell model for the open quantum mechanical nuclear system, Rep. Prog. Phys. \textbf{54}, 635 (1991).
\bibitem{open_3} H. Carmichael, {\it An Open System Approach to Quantum Optics} (Springer-Verlag, Heidelberg, 1993).
\bibitem{open_4} F. M. Dittes, The decay of quantum systems with a small number of open channels, Phys. Rev. \textbf{339}, 215 (2000).
\bibitem{open_5} A. J. Daley, Quantum trajectories and open many-body quantum systems, Adv. Phys. \textbf{63}, 77 (2014).
% effective non-Hermitian Hamiltonian
\bibitem{eff_1} V. V. Konotop, J. Wang, and D. A. Zezyulin, Nonlinear waves in $\mathcal{PT}$-symmetric systems, Rev. Mod. Phys. \textbf{88}, 035002 (2016).
\bibitem{eff_2} Y. Ashida, S. Furukawa, and M. Ueda, Quantum critical behavior influenced by measurement backaction in ultracold gases, Phys. Rev. A \textbf{94}, 053615 (2016).
\bibitem{eff_3} Y. Ashida, S. Furukawa, and M. Ueda, Parity-time-symmetric quantum critical phenomena, Nat. Commun. \textbf{8}, 15791 (2017).
\bibitem{eff_4} R. El-Ganainy, K. G. Makris, M. Khajavikhan, Z. H. Musslimani, S. Rotter, and D. N. Christodoulides, Non- Hermitian physics and PT symmetry, Nat. Phys. \textbf{14}, 11 (2018).
\bibitem{eff_5}  L. Jin and Z. Song, Bulk-boundary correspondence in a non-hermitian system in one dimension with chiral inversion symmetry, Phys. Rev. B \textbf{99}, 081103(R) (2019).
% Bender and Boettcher PT seminal work
\bibitem{PT_seminal} C. M. Bender and S. Boettcher, Phys. Rev. Lett. \textbf{80}, 5243 (1998).

% real-complex transition without PT-symmetry
\bibitem{except_1} R. Hamazaki, K. Kawabata, and M. Ueda, Non-Hermitian Many-body Localization, Phys. Rev. Lett. \textbf{123}, 090603 (2019).
\bibitem{except_2} T. Liu, S. Cheng, H. Guo, and G. Xianlong, Fate of Majorana zero modes, exact location of critical states, and unconventional real-complex transition in non-Hermitian quasiperiodic lattices, Phys. Rev. B \textbf{103}, 104203 (2021).

% quantum field theory and mathematical physics
\bibitem{PT_seminal_1} C. M. Bender, S. Boettcher, and P. N. Meisinger, PT-symmetric quantum mechanics, J. Math. Phys. (NY) \textbf{40}, 2201 (1999).
\bibitem{PT_seminal_2} C. M. Bender, D. C. Brody, and H. F. Jones, Complex Extension of Quantum Mechanics, Phys. Rev. Lett. \textbf{89}, 270401 (2002).
\bibitem{PT_seminal_3} C. M. Bender, Making sense of non-Hermitian Hamiltonians, Rep. Prog. Phys. \textbf{70}, 947 (2007).
\bibitem{PT_seminal_4} C. M. Bender, D. C. Brody, and M. P. M\"{u}ller, Hamiltonian for the Zeros of the Riemann Zeta Functions, Phys. Rev. Lett. \textbf{118}, 130201 (2017).

%condensed matter physics
\bibitem{cmp_1} O. Bendix, R. Fleischmann, T. Kottos, and B. Shapiro, Exponentially Fragile $\mathcal{PT}$ Symmetry in Lattices with Localized Eigenmodes, Phys. Rev. Lett. \textbf{103}, 030402 (2009).
\bibitem{cmp_2} L. Jin and Z. Song, Solutions of PT -symmetric tightbinding chain and its equivalent hermitian counterpart, Phys. Rev. A \textbf{80}, 052107 (2009).

% optical systems
\bibitem{opt_1} R. El-Ganainy, K. G. Makris, D. N. Christodoulides, and Z. H. Musslimani, Theory of coupled optical PT-symmetric structures, Opt. Lett. \textbf{32}, 2632 (2007).
\bibitem{opt_2} K. G. Makris, R. El-Ganainy, D. N. Christodoulides, and Z. H. Musslimani, Beam Dynamics in $\mathcal{PT}$ Symmetric Optical Lattices, Phys. Rev. Lett. \textbf{100}, 103904 (2008).
\bibitem{opt_3} Z. H. Musslimani, K. G. Makris, R. El-Ganainy, and D. N. Christodoulides, Optical Solitons in $\mathcal{PT}$ Periodic Potentials, Phys. Rev. Lett. \textbf{100}, 030402 (2008).
\bibitem{opt_4} S. Longhi, Bloch Oscillations in Complex Crystals with $\mathcal{PT}$ Symmetry, Phys. Rev. Lett. \textbf{103}, 123601 (2009).
\bibitem{opt_5} S. Longhi, Dynamic localization and transport in complex crystals, Phys. Rev. B \textbf{80}, 235102 (2009).
\bibitem{opt_6} A. Guo, G. J. Salamo, D. Duchesne, R. Morandotti, M. VolatierRavat, V. Aimez, G. A. Siviloglou, and D. N. Christodoulides, Observation of $\mathcal{PT}$-Symmetry Breaking in Complex Optical Potentials, Phys. Rev. Lett. \textbf{103}, 093902 (2009).

% control  gain and loss
\bibitem{cgl_1}  C. E. R\"{u}ter, K. G. Makris, R. El-Ganainy, D. N. Christodoulides, M. Segev, and D. Kip, Observation of parity-time symmetry in optics, Nat. Phys. \textbf{6}, 192 (2010).
\bibitem{cgl_2} A. Regensburger, C. Bersch, M.-A. Miri, G. Onishchukov, D. N. Christodoulides, and U. Peschel, Parity-time synthetic photonic lattices, Nature (London) \textbf{488}, 167 (2012).
\bibitem{cgl_3} B. Peng, S. K. \"{O}zdemir, F. Lei, F. Monifi, M. Gianfreda, G. L. Long, S. Fan, F. Nori, C. M. Bender, and L. Yang, Parity-time- symmetric whispering-gallery microcavities, Nat. phys. \textbf{10}, 394 (2014).
\bibitem{cgl_4} L. Feng, Z. J. Wong, R.-M. Ma, Y. Wang, and X. Zhang, Single-mode laser by parity-time symmetry breaking, Science \textbf{346}, 972 (2014).
\bibitem{cgl_5} H. Hodaei, M.-A. Miri, M. Heinrich, D. N. Christodoulides, and M. Khajavikhan, Parity-time-symmetric microring lasers, Science \textbf{346}, 975 (2014).
\bibitem{cgl_6} B. Zhen, C. W. Hsu, Y. Igarashi, L. Lu, I. Kaminer, A. Pick, S.-L. Chua, J. D. Joannopoulos, and M. Solja\v{c}\'{e}, Spawning rings of exceptional points out of dirac cones, Nature (London) \textbf{525}, 354 (2015).
\bibitem{cgl_7} J. M. Zeuner, M. C. Rechtsman, Y. Plotnik, Y. Lumer, S. Nolte, M. S. Rudner, M. Segev, and A. Szameit, Observation of a Topological Transition in the Bulk of a Non-Hermitian System, Phys. Rev. Lett. \textbf{115}, 040402 (2015).
\bibitem{cgl_8} C. Poli, M. Bellec, U. Kuhl, F. Mortessagne, and H. Schomerus, Selective enhancement of topologically induced interface states in a dielectric resonator chain, Nat. Commun. \textbf{6}, 6710 (2015).
\bibitem{cgl_9} J. Doppler, A. A. Mailybaev, J. B\"{o}hm, U. Kuhl, A. Girschik, F. Libisch, T. J. Milburn, P. Rabl, N. Moiseyev, and S. Rotter, Dynamically encircling an exceptional point for asymmetric mode switching, Nature (London) \textbf{537}, 76 (2016).

% Anderson localization
\bibitem{Anderson}P. W. Anderson, Absence of diffusion in certain random lattices, Phys. Rev. {\bf 109}, 1492 (1958).

% AA moodel
\bibitem{AA} S. Aubry and G. Andr\'{e},  Analyticity breaking and anderson localization in incommensurate lattices, Ann. Isr. Phys. Soc. \textbf{3}, 18 (1980).

% realize the AA model
\bibitem{exp_AA} G. Roati, C. D'Errico, L. Fallani, M. Fattori, C. Fort, M. Zaccanti, G. Modugno, M. Modugno, and M. Inguscio, Anderson localization of a non-interacting boseCeinstein condensate, Nature (London) \textbf{453}, 895 (2008).
\bibitem{Laurent1} L. Sanchez-Palencia, D. Cl\'ement, P. Lugan, P. Bouyer, G. V. Shlyapnikov, and A. Aspect, Anderson localization of expanding Bose-Einstein condensates in random potentials, Phys. Rev. Lett. \textbf{98}, 210401 (2007).
\bibitem{Laurent2} L. Sanchez-Palencia and M. Lewenstein, Disordered quantum gases under control, Nat. Phys. \textbf{6}, 87 (2010).
\bibitem{Laurent3} F. Jendrzejewski, A. Bernard, K. M\"{u}ller, P. Cheinet, V.Josse, M. Piraud, L. Pezz\'e, L. Sanchez-Palencia, A. Aspect, and P. Bouyer, Three-dimensional localization of ultracold atoms in an optical disordered potential, Nat. Phys. \textbf{8}, 398 (2012).
\bibitem{Laurent4} L. Sanchez-Palencia, Ultracold gases: At the edge of mobility, Nat. Phys. \textbf{11}, 525 (2015).%
 \bibitem{Laurent5} H. Yao, H. Khouldi, L. Bresque, and L. Sanchez-Palencia, Critical behavior and fractality in shallow one-dimensional quasiperiodic potentials, Phys. Rev. Lett. \textbf{123}, 070405 (2019). %
\bibitem{Laurent6} H. Yao, T. Giamarchi, and L. Sanchez-Palencia, Lieb-Liniger bosons in a shallow quasiperiodic potential: Bose glass phase and fractal Mott Lobes, Phys. Rev. Lett. \textbf{125}, 060401 (2020). %
\bibitem{Billy} J. Billy, V. Josse, Z. Zuo, A. Bernard, B. Hambrecht, P. Lugan, D. Cl\'ement, L. Sanchez-Palencia, P. Bouyer, and A. Aspect, Direct observation of Anderson localization of matter waves in a controlled disorder, Nature (London) \textbf{453}, 891 (2008).

% non-Hermitian Hatano-Nelson model
\bibitem{HN_1} N. Hatano and D. R. Nelson, Localization Transitions in non-Hermitian Quantum Mechanics, Phys. Rev. Lett. \textbf{77}, 570 (1996).
\bibitem{HN_2} N. Hatano and D. R. Nelson, Vortex pinning and non-Hermitian quantum mechanics, Phys. Rev. B \textbf{56}, 8651 (1997).
\bibitem{HN_3} N. Hatano and D. R. Nelson, Non-Hermitian delocalization and eigenfunctions, Phys. Rev. B \textbf{58}, 8384 (1998).
% non-Hermitian AA model
\bibitem{NH_AA_2} C. Yuce, $\mathcal{PT}$-symmetric Aubry-Andr\'{e} model, Phys. Lett. A \textbf{378}, 2024 (2014).
\bibitem{NH_AA_3} Q.-B. Zeng, S. Chen, and R. L\"{u}, Anderson localization in the non-Hermitian Aubry-Andre-Harper model with physical gain and loss, Phys. Rev. A \textbf{95}, 062118 (2017).
\bibitem{NH_AA_4} Q.-B. Zeng, Y.-B. Yang, and Y. Xu, Topological phases in non-Hermitian Aubry-Andr\'{e}-Harper models, Phys. Rev. B \textbf{101}, 020201 (2020).
\bibitem{Longhi1} S. Longhi, Topological Phase Transition in non-Hermitian Quasicrystals, Phys. Rev. Lett. {\bf 122}, 237601 (2019).
\bibitem{Longhi2} S. Longhi, Metal-insulator phase transition in a non-Hermitian Aubry-Andr\'{e}-Harper Model, Phys. Rev. B {\bf 100}, 125157 (2019).
% topological nature of the localization transition in non-Hermitian Hatano-Nelson model
\bibitem{HN_5} Z. Gong, Y. Ashida, K. Kawabata, K. Takasan, S. Higashikawa, and M. Ueda, Topological Phases of Non-Hermitian Systems, Phys. Rev. X \textbf{8}, 030079 (2018).
% PT-symmetry protected Anderson Localization phase
\bibitem{PT_AA} S. Schiffer, X.-J. Liu, H. Hu, and J. Wang, Anderson localization in a robust $\mathcal{PT}$-symmetric phase of a generalized Aubry-Andr\'{e} model, Phys. Rev. A \textbf{103}, L011302 (2021).

% Mott uncovers and proposes the mobility edges
\bibitem{mott} N. Mott, The mobility edge since 1967, J. Phys. C  \textbf{20}, 3075 (1987).


% slow-varying
\bibitem{slow_1} S. Das. Sarma, S. He, and X. C. Xie, Mobility Edge in a Model One-Dimensional Potential, Phys. Rev. Lett. \textbf{61}, 2144 (1988).
\bibitem{slow_2} S. Das. Sarma, S. He, and X. C. Xie, Localization, mobility edges, and metal-insulator transition in a class of one-dimensional slowly varying deterministic potentials, Phys. Rev. B \textbf{41}, 5544 (1990).
% off-diagonal disorder
\bibitem{off_1} T. Liu, G. Xianlong, S. Chen, and H. Gao, Localization and mobility edges in the off-diagonal quasiperiodic model with slowly varying potentials, Phys. Lett. A \textbf{381}, 3683 (2017).
\bibitem{off_2} T. Liu and H. Guo, Mobility edges in off-diagonal disordered tight-binding models, Phys. Rev. B \textbf{98}, 104201 (2018).
% long-range hoppings
\bibitem{long-range} J. Biddle and S. Das Sarma, Predicted mobility edges in one-dimensional incommensurate optical lattices: An exactly solvable model of Anderson localization, Phys. Rev. Lett. \textbf{104}, 070601 (2010).
% general quasiperiodic potential
\bibitem{general_quasi_1} S. Ganeshan, J. H. Pixley, S. Das Sarma, Nearest Neighbor Tight Binding Models with an Exact Mobility Edge in One Dimension, Phys. Rev. Lett. \textbf{114}, 146601 (2015).
\bibitem{general_quasi_2} Z. Xu, H. Huangfu, Y. Zhang, S. Chen, Dynamical observation of mobility edges in one-dimensional incommensurate optical lattices, New. J. Phys. \textbf{22}, 013036 (2020).
\bibitem{general_quasi_3} Y. Wang, X. Xia, Y. Wang, Z. Zheng, and X.-J. Liu, Duality between two generalized Aubry-Andr\'{e} models with exact mobility edges, arXiv: 2012.09756 (2021).

% MEs in experiments
%\bibitem{exp_1} F. Jendrzejewski, A. Bernard, K. Mueller, P. Cheinet, V. Josse, M. Piraud, L. Pezz\'{e}, L. Sanchez-Palencia, A. Aspect, and P. Bouyer, Three-dimensional localization of ul- tracold atoms in an potical disordered potential, Nature Physics \textbf{8}, 398 (2012).
%\bibitem{exp_2} W. McGehee, S. Kondov, W. Xu, J. Zirbel, and B. DeMarco, Three-dimensional Anderson localization in vari- able scale disorder, Phys. Rev. Lett. \textbf{111}, 145303 (2013).
%\bibitem{exp_3} G. Semeghini, M. Landini, P. Castilho, S. Roy, G. Spagnolli, A. Trenkwalder, M. Fattori, M. Inguscio, and G. Modugno, Measurement of the mobility edge for 3D An- derson localization, Nature Physics \textbf{11}, 554 (2015).
%\bibitem{exp_4} H. P. L\"{u}schen, S. Scherg, T. Kohlert, M. Schreiber, P. Bordia, X. Li, S. D. Sarma, and I. Bloch, Single-particle mobility edge in a one-dimensional quasiperiodic optical lattice, Phys. Rev. Lett. \textbf{120}, 160404 (2018).
%\bibitem{exp_5} F. A. An, E. J. Meier, and B. Gadway, Engineering a flux- dependent mobility edge in disordered zigzag chains, Phys. Rev. X \textbf{8}, 031045 (2018).
%\bibitem{exp_6} T. Kohlert, S. Scherg, X. Li, H. P. L\"{u}schen, S. D. Sarma, I. Bloch, and M. Aidelsburger, Observation of many-body localization in a one-dimensional system with singleparticle mobility edge, Phys. Rev. Lett. \textbf{122}, 170403 (2019).
%\bibitem{exp_7} F. A. An, K. Padavic, E. J. Meier, S. Hegde, S. Ganeshan, J. H. Pixley, S. Vishveshwara, B. Gadway, Observa- tion of tunable mobility edges in generalized Aubry-Andr\'{e} lattices, arXiv: 2007.01393.



% PT symmetry and mobility edges
\bibitem{YLiu} Y. Liu, X.-P. Jiang, J. Cao, and S. Chen, Non-Hermitian mobility edges in one-dimensional quasicrystals with parity-time symmetry, Phys. Rev. B \textbf{101}, 174205 (2020).
\bibitem{QZeng} Q.-B. Zeng and Y. Xu, Winding numbers and generalized mobility edges in non-Hermitian systems, Phys. Rev. Research \textbf{2}, 033052 (2020).
\bibitem{TLiu} T. Liu, H. Guo, Y. Pu, and S. Longhi, Generalized Aubry-Andr\'{e} self-duality and mobility edges in non-Hermitian quasiperiodic lattices, Phys. Rev. B \textbf{102}, 024205 (2020).

\bibitem{Flach1} S. Flach, D. Leykam, J. D. Bodyfelt, P. Matthies, and A. S. Desyatnikov, Detangling flat bands into Fano lattices, Europhys. Lett. {\bf 105}, 30001 (2014).
\bibitem{Flach2} W. Maimaiti, A. Andreanov, H. C. Park, O. Gendelman, and S. Flach, Compact localized states and flat-band generators in one dimension, Phys. Rev. B {\bf 95}, 115135 (2017).
\bibitem{Flach3} A. Ramachandran, A. Andreanov, and S. Flach, Chiral flat bands: Existence, engineering, and stability, Phys. Rev. B {\bf 96}, 161104(R) (2017).
%\bibitem{Flach4} J. D. Bodyfelt, D. Leykam, C. Danieli, X. Yu, and S. Flach, Flatbands under correlated perturbations, Phys. Rev. Lett. {\bf 113}, 236403 (2014).
\bibitem{Flach5} R. Khomeriki and S. Flach, Landau-Zener Bloch oscillations with perturbed flat bands, Phys. Rev. Lett. {\bf 116}, 245301 (2016).
\bibitem{Gneiting} C. Gneiting, Z. Li, and F. Nori, Lifetime of flatband states, Phys. Rev. B {\bf 98}, 134203 (2018).

\bibitem{LIU} Z. Liu, E. J. Bergholtz, H. Fan, and A. M. L\"{a}uchli, Fractional Chern insulators in topological flat bands with higher Chern number, Phys. Rev. Lett. {\bf 109}, 186805 (2012).

\bibitem{Bodyfelt} D. Bodyfelt, D. Leykam, C. Danieli, X. Yu, and S. Flach, Flatbands under correlated perturbations, Phys. Rev. Lett. {\bf 113}, 236403 (2014).
\bibitem{Danieli} C. Danieli, J. D. Bodyfelt, and S. Flach, Flat-band engineering of mobility edges, Phys. Rev. B {\bf 91}, 235134 (2015).
\bibitem{Jazaeri} A. Jazaeri and I. I. Satija, Localization transition in incommensurate non-Hermitian systems, Phys. Rev. E {\bf 63}, 036222 (2001).
\bibitem{IPR} D. J. Thouless, A relation between the density of states and range of localization for one dimensional random systems, J. Phys. C: Solid State Phys. {\bf 5}, 77 (1973).
\end{thebibliography}
\end{document}